\documentstyle[aps,preprint]{revtex}
\begin{document}
\tighten
\title{\bf
Spectroscopy of the quantum black hole}
\author{Jacob D. Bekenstein\footnote[1]{ Electronic
mail: bekenste@vms.huji.ac.il}}
\address{\it Racah Institute of
Physics, Hebrew University of Jerusalem, Givat Ram, Jerusalem 91904,
Israel}
\author{V. F. Mukhanov\footnote[2]{Electronic mail:
mukhanov@itp.phys.ethz.ch}}   \address{\it Institute of Theoretical Physics,
ETH-Hoenggerberg, CH-8093 Z\"urich, Switzerland}
\date{Received \today}
\maketitle \begin{abstract}
We develop the idea that, in quantum gravity where the horizon
fluctuates, a black hole should have a discrete mass spectrum with concomitant
line emission. Simple arguments fix the spacing of the lines, which should be
broad but unblended.  Assuming uniformity of the matrix elements for quantum
transitions between near levels, we work out the probabilities for
the emission of a specified series of quanta and the intensities of the
spectral lines.  The thermal character of the radiation is
entirely due to the degeneracy of the levels, the same degeneracy that
becomes manifest as black hole entropy.  One prediction is that there should
be no lines with wavelength of order the black hole size or larger.  This
makes it possible to test quantum gravity with black holes well above Planck
scale.

\end{abstract}  

In Hawking's conception\cite{Hawk} of the origin of black hole
radiance, modes prepared in the {\it pure\/} vacuum state
at ${\cal I_{-}}$ contract, pass through the center of the collapsing star,
and make it out to ${\cal I_{+}}$  after spending a long time ``skimming the
horizon''.  The mode frequenies are redshifted down and the modes become
populated by real quanta, the partners of those other members of the created
pairs that have fallen through the horizon.  The
outgoing quanta have a thermal  spectrum  and
statistics\cite{statistics} that bear the imprint of passage
through the potential barrier surrounding the black hole.  The universality of
this thermal character is viewed as due to the long sojourn of the emergent
quanta close to the horizon.

Quantum fluctuations of the horizon, which are expected within any
reasonable theory of quantum gravity, must change this simple picture.  The
Hawking quanta will not be able to ``hover'' at a nearly fixed distance from
the horizon since the latter's very location has to fluctuate.  Thus one
suspects a modification of the character of the radiation when quantum gravity
effects are properly taken into account, even for black holes very massive
with respect to the Planck scale.  In this letter we consider what
modification of the naive  Hawking radiance spectrum and statistics might
reasonably be expected from quantum gravity.

Quantum systems of finite size more often than not display a discrete energy
spectrum.  Since the dynamics of a black hole responsible for its unique
character refer to the finite region enclosed by the horizon, one may thus
expect the mass spectrum of a black hole to display discreteness. Long ago the
case was made \cite{Bek,Mukh} that the black hole's  horizon area  should be
quantized in integers. In particular (we use units with $G=c=1$)
\begin{equation}  A = \alpha\hbar n;  \qquad n=1, 2\dots   \label{area}
\end{equation}  where $\alpha$ is a pure number.  This type of quantization
law has since been considered by many workers \cite{Many,DanSchiff}.  In
particular, it is in agreement with the idea that area should be quantized in
canonical quantum gravity \cite{Ashtekar}.  The rule (\ref{area}) implies that
the mass of a nonrotating neutral black hole has a discrete spectrum.  In
general an energy level labeled by $n$ will be degenerate; let its
multiplicity be denoted by $g(n)$.   Further, it is natural to identify the
entropy $S_{BH}={1\over 4}A/\hbar+{\rm const.}$ of the black hole in the level
$n$ with $\ln g(n)$.   The  rule (\ref{area}) in conjunction with the natural
assumption that $g(1)=1$ or $S_{\rm BH}(n=1)=0$ (nondegenerate ground state)
then forces us too choose $g(n)=e^{\alpha (n-1)/4}$.  But $g(n)$ must be
integral, so $\alpha=4\ln k$ where $k=2, 3, 4\dots$.

One of us \cite{Mukh} has argued that the choice $k=2$ is preferred because it
would make the spacing in entropy of consecutive energy levels exactly one
bit, an attractive value from the information--theoretic point of view.  One
convincing piece of evidence in favor of this choice is that it leads to
$g(n)=2^{n-1}$ whereas the number of ways in which the black hole at level $n$
can be built starting from a situation with $A=0$ (no black hole) and going up
the staircase of levels by various combinations of steps is also $2^{n-1}$.
Thus for $k=2$ the degeneracy actually quantifyies the number of ways the
black hole in the particular level could have formed. Put another way, the
decay from the $n$--th level to the no--black hole state in such scheme can be
accomplished in exactly $2^{n-1}$ sequences of steps \cite{DanSchiff}.  There
is thus weighty evidence in  favor of $k=2$ or $\alpha=4\ln 2$.

With this choice the energy spacing between consecutive levels for $M\gg
\hbar$ corresponds to the (fundamental) frequency
\begin{equation}
\varpi={\ln 2\over 8\pi M}
\label{basicfreq}
\end{equation}
Clearly the radiation emitted as the black hole decays will be concentrated
in lines at integer multiples of the fundamental frequency $\varpi$.  Even if
some broadening occurs or if the spectrum is blurred by two--quanta emission
per jump, it is clear that the spectrum will be radically
different from the one entertained in the standard discussion of Hawking
radiance.  One consequence of the physics here considered is that no
radiation (or very little) should be emitted below $\varpi$.
Observation of this type of spectrum for any black hole would immediately make
quantum gravity effects observable well above the Planck scale.   Herein
lies the importance of our considerations.

Information about the intensity of the lines emitted by the black hole in the
various particles available, and the statistics of these emissions can
be obtained by first focusing on the de--excitation probabilities.  We define
these as follows.  Describe the decay of the black hole during any interval of
observer time $\Delta t$ by giving a sequence of integers $\{n_1, n_2, \cdots,
n_j\}$.  This means that during  $\Delta t$ the black hole first jumped down
$n_1$ elementary levels in one go, then $n_2$ levels, $\cdots$, and concluded
by jumping down $n_j$ levels.  In the process it emitted a quantum of some
species of energy $n_1\hbar\varpi$, then a quantum of  energy
$n_2\hbar\varpi$, and so on.

These sequences can occur with any length $j$.  The sequence of zero
length, formally written $\{0\}$, represents the eventuality that the black
hole did not decay at all in the interval $\Delta t$.  One can associate with
any sequence of length $j$ a {\it conditional\/} probability $P_{\Delta
t}(\{n_1, n_2, \cdots, n_j\}|j)$ whose dependence on the species of quanta we
ignore in this preliminary exploration.  Obviously normalization requires
that   \begin{equation} \sum_{\{n_1, n_2, \cdots , n_j\}} P_{\Delta t}(\{n_1,
n_2, \cdots , n_j\}|j)=1, \label{normalization} \end{equation} where the all
the $n_k$ are nonvanishing.

Even before dealing with the conditional probability  $P_{\Delta t}(\{n_1,
n_2, \cdots , n_j\}|j)$ given that the hole decayed by a sequence of $j$
jumps, we must evaluate the probability $p_{\Delta t}(j)$ for a sequence of
exactly $j$ jumps (and $j$ emitted quanta) in time $\Delta t$.  If $\Delta t$
is not vastly large, the black hole will have changed very little after time
$\Delta t$ in terms of its emission properties (it will have nearly the same
mass).  Then the probabilities  $p_{\Delta t}(j)$ for the next time interval
$\Delta t$ can be taken as equal to the previous $p_{\Delta t}(j)$.  On the
basis of this and the fact that  one does not attach significance to a {\it
failure\/} of the hole to jump at some point in a sequence, we can write, for
instance, \begin{equation} p_{2\Delta t}(1) = p_{\Delta t}(0) p_{\Delta t}(1)
+ p_{\Delta t}(1)p_{\Delta t}(0) =2 p_{\Delta t}(0) p_{\Delta t}(1)
\label{onejump} \end{equation} In words: a jump by one level in the interval
$2\Delta t$ can take place either by the jump taking place in the first half
of the interval with no jump in the second half, or by no jump taking place in
the first half with the jump by one step taking place in the second half. The
above equality may be generalized to any odd $j$ by \begin{equation}
p_{2\Delta t}(j) = 2 p_{\Delta t}(0) p_{\Delta t}(j) + 2 p_{\Delta
t}(1)p_{\Delta t}(j-1) + \cdots + 2 p_{\Delta t}(j/2+1/2) p_{\Delta
t}(j/2-1/2)
\label{oddnumber} \end{equation}
and to any even $j$ by
\begin{equation}
p_{2\Delta t}(j) = 2 p_{\Delta t}(0) p_{\Delta t}(j) + 2 p_{\Delta
t}(1)p_{\Delta t}(j-1) + \cdots + [ p_{\Delta t}(j/2)]^2
\label{evennumber}
\end{equation}

These equations may be supplemented by the very obvious one
\begin{equation}
p_{2\Delta t}(0) = [p_{\Delta t}(0)]^2
\label{zeronumber}
\end{equation}
which just says that the survival probability of the black hole in a given
level over time $2\Delta t$ is product of the survival probability over time
$\Delta t$ by itself.  The solution of Eq.~(\ref{zeronumber}) is
\begin{equation}
p_{\Delta t}(0) = e^{-\Delta t/\tau}
\label{p0}
\end{equation}
where $\tau$ is a survival timescale to be determined.  Substituting
Eq.~(\ref{p0}) into Eq.~(\ref{onejump}) gives a simple functional equation for
$p_{\Delta t}(1)$ whose general solution is  \begin{equation} p_{\Delta t}(1)
=(\Delta t/\tau^*)\, e^{-\Delta t/\tau} \label{p1} \end{equation}
where $\tau^*$ is a positive constant---possibly distinct from $\tau$---to be
determined.  Now substituting Eqs.~(\ref{p0})--(\ref{p1}) into
Eq.~(\ref{evennumber}) gives a functional equation for $p_{\Delta t}(2)$ which
we solve by
\begin{equation}
p_{\Delta t}(2) =(1/2)(\Delta t/\tau^*)^2\, e^{-\Delta t/\tau}
\label{p2}
\end{equation}

On the basis of Eqs.~(\ref{p0})--(\ref{p2}) let us guess that for any $j$
\begin{equation}
p_{\Delta t}(j) =(1/j!)(\Delta t/\tau^*)^j\, e^{-\Delta t/\tau}
\label{pj}
\end{equation}
To prove this guess for any odd $j$ we substitute Eq.~(\ref{p0}) and our guess
in Eq.~(\ref{oddnumber}), factor out the common factor $(\Delta t/\tau^*)^j
e^{-j\Delta t/\tau}$, and multiply the resulting equation by $j!$ to get
 \begin{equation}
2^{j-1}-1 = j + {j(j-1)\over 2!} +{j(j-1)(j-2)\over 3!} +\cdots +{j(j-1)\cdots
(j/2+1/2)\over 2 (j/2)!}
 \label{check}
\end{equation}
But this is plainly correct by the binomial expansion of $(1+1)^j$.
Thus Eq.~(\ref{pj}) is correct for all odd $j$.  The proof of Eq.~(\ref{pj})
for even $j$ proceeds similarly from Eq.~(\ref{evennumber}).  We must
now check the normalization of $p_{\Delta t}(j)$:
\begin{equation}
\sum_j p_{\Delta t}(j) = \exp(\Delta t/\tau^* - \Delta t/\tau)
 \label{norm}
\end{equation}
Obviously we must set $\tau^*=\tau$ to get normalization.

Thus we have shown that the {\it lengths\/} of the sequence $\{n_1, n_2,
\cdots , n_j\}$ (in other words, the number of quanta emitted during $\Delta
t$) follows a Poisson probability distribution, [Eq.~(\ref{pj}) with
$\tau^*=\tau$]  depending on a single parameter $\tau$.  It is a consequence
of Eq.~(\ref{pj}) that the mean value of $j$ is simply $\langle
j\rangle =\Delta t/\tau$.

Let us now return to the discussion preceding the introduction of
$p_{\Delta t}(j)$.  Given that the black hole has decayed during the interval
$\Delta t$ by a sequence  $\{n_1, n_2, \cdots , n_j\}$ of length $j$, what is
the corresponding conditional probability distribution $P_{\Delta t}(\{n_1,
n_2, \cdots , n_j\}|j)$  ?   For $j=1$ the answer is not difficult to guess.
When the black hole jumps $n_1$ steps down the ladder of levels, the
degeneracy $g(n)$ changes  by a factor of $2^{-n_1}$.   If we assume the
``matrix element'' for a jump is the same for all $n_1\not = 0$, then the
probability for the jump must be proportional to the final level's
degeneracy.  Thus \begin{equation} P_{\Delta t}(\{n_1\}|1)\propto 2^{-n_1}
\label{jump}
\end{equation}
(The case $n_1=0$ is not included here because it belongs to $j=0$). From the
normalization $\sum_{n_1=1}^\infty P_{\Delta t}(\{n_1\}|1)=1$ we see
immediately that the proportionality constant in Eq.~(\ref{jump}) is unity.
Combining Eq.~(\ref{jump}) with  Eq.~(\ref{p0}) we get for the {\it a
priori\/} probability that  the hole jumps once $n_1$ steps in $\Delta t$
\begin{equation}
P_{\Delta t}(\{n_1\})= P_{\Delta t}(\{n_1\}|1)\, p_{\Delta t}(1) =
(\Delta t/2^{n_1}\tau)\, e^{-\Delta t/\tau}. \label{jump1}
\end{equation}

Now consider the case $j > 1$.  Up to now we have not regarded $\Delta t/\tau$
as having any particular value.  However, it is clear from Eq.~(\ref{pj}) that
when $\Delta t\ll\tau$, $p_{\Delta t}(j)$ is very small for $j>1$.  We are
thus interested in calculating $P_{\Delta t}(\{n_1, n_2,
\cdots , n_j\}|j)$ only for $\Delta t/\tau$ not small compared to unity.
However, it is a useful strategy to imagine the interval $\Delta t$ divided
into a large number $N\gg j$ of subintervals of equal durations
$\epsilon\equiv\Delta t/N$.  Because quantum transitions ``do not take time'',
we can think of the $j$ jumps as taking place each in one of $j$ of the $N$
subintervals.  There is no need to consider two or more jumps in one
subinterval because, by our earlier remark, the probability of such an event
would be very small because $N$ is large.

Now the $j$ ``active'' subintervals can be chosen out of the $N$ subintervals
in a total of $N!/j!(N-j)!\approx N^j/j!$ ways, where the approximation is a
good one because $j\ll N$.  Thus the analog of Eq.~(\ref{jump1}) is
\begin{equation}
P_{\Delta t}(\{n_1, n_2, \cdots , n_j\})\approx
(N^j/j!)P_\epsilon(\{n_1\}|1)P_\epsilon(\{n_2\}|1)\cdots P_\epsilon(\{n_j\}|1)
[p_\epsilon(1)]^j [p_\epsilon(0)]^{N-j} \label{jumpj} \end{equation}
where the last factor is the probability for the $N-j$ subintervals during
which the hole does not decay.  Substituting from Eq.~(\ref{pj}) and replacing
$\epsilon\Rightarrow\Delta t/N$ we have
\begin{equation}
P_{\Delta t}(\{n_1, n_2, \cdots , n_j\})= (1/j!)\,2^{-(n_1+n_2\, +\cdots\, +
n_j)}\,(\Delta t/\tau)^j\, e^{-\Delta t/\tau} \label{final}
\end{equation}
Under the stated assumptions this equation should be exact as
$N\rightarrow\infty$.  The set of probabilities is normalized in the sense
that when summed for fixed $j$ over all configurations $\{n_1,
n_2,\cdots,n_j\}$ with every $n_i$ starting from one, the result is exactly
$p_{\Delta t}(j)$, so that the further sum over $j$ yields unity.  Another
thing to notice is that
\begin{equation}
P_{\Delta t}(\{n_1, n_2, \cdots , n_j\}|j)\equiv
P_{\Delta t}(\{n_1, n_2, \cdots , n_j\})/p_{\Delta t}(j)=2^{-(n_1+n_2 +\cdots
+ n_j)}. \label{conditional}
\end{equation}
Thus given that the black hole jumped $j$ times in $\Delta t$,
the distribution of the sizes of the jumps is independent of $\Delta t$ as
well as any scale set by the black hole.  (In fact, we could obtain
Eq.~(\ref{conditional}) directly by repeating our argument that if the
transition  matrix element is constant, the probabilities depend only on the
multiplicity of the final level.)  One consequence of this is that the
spectrum and the statistics of the radiation are steady so long as the mass of
the hole changes by little.

Let us now estimate the parameter $\tau$.  The conditional probabilities
Eq.~(\ref{conditional}) are normalized.  Since $\langle n_i\rangle
=\sum_{n_i=1}^\infty n_i 2^{-n_i}=2$, each one of the $j$ quanta  carries
energy $2\hbar\varpi$ on average.  Since the mean value of $j$ is $\Delta
t/\tau$, we see that the mean decrease in black hole mass during $\Delta t$ is
just $2\hbar\varpi\Delta t/\tau$ so that  \begin{equation}
d\langle M\rangle/dt =- 2\hbar\varpi\Delta t/\tau
\end{equation}
Due to the thermal character of the radiation, we may
estimate $d\langle M\rangle/dt$ by using the Stefan--Boltzmann law for a
surface of area $16\pi \langle M\rangle^2$ at temperature
$(\hbar/8\pi)\langle M\rangle^{-1}$: \begin{equation}
{d\langle M\rangle\over dt} =- {\gamma\,\hbar\over 1530\pi \langle M\rangle^2}
\end{equation}
Here $\gamma$ accounts for two fact: the radiation comes out only in
lines and not over the entire black--body spectrum, and the emission is by a
variety of particles.  The first fact tends to force $\gamma$ under
unity, while the second promotes a value well above unity.  On the
balance we feel $\gamma$ should be regarded as being unity to within
an order of magnitude. Inserting the value of $\varpi$ and comparing these
equations gives us \begin{equation} \tau=3840\gamma^{-1}\langle M\rangle\ln
2 \end{equation} It is plain that $\tau$, the mean time between quantum leaps,
is large compared to the dynamical black hole timescale $\langle M\rangle$.

We now calculate the probability distribution $p_{\Delta t}(k|n_k\varpi)$ that
in the course of time $\Delta t$, the black hole emits $k$ quanta of
frequency $n_k\varpi$.  If the hole makes $j$ jumps in $\Delta t$, the number
of ways of selecting the $k$ quanta out of the $j$ is $C^j_k\equiv
j![k!(j-k)!]^{-1}$, and the probability of each such selection is
$2^{-(n_1+n_2 +\cdots + n_{j-k})}(2^{-n_k})^k$ with none of the $n_1, n_2,
\cdots\,, n_{j-k}$ agreeing with $n_k$.  We must sum the product over all
allowed values of $n_1\not= n_k$,  $n_2\not= n_k$, {\it etc.\/}.  Since
$\sum_{n=1}^\infty 2^{-n}=1$, the sum over each of the $j-k$ factors
$2^{-n_i}$  gives a factor $1-2^{-n_k}$.  Thus \begin{equation}
p_{\Delta t}(k|n_k\varpi, j)={j!\over k!(j-k)!}\left(1-{1\over
2^{n_k}}\right)^j\left({1\over 2^{n_k}-1}\right)^k  \label{intermediate}
\end{equation}
We now multiply this conditional probability by the absolute probability
distribution $p_{\Delta t}(j)$ in Eq.~(\ref{pj}), and sum over all
$j \ge k$.  Since $\sum_{j=s}^\infty x^j/(j-s)!=e^x\,x^s$,
we have again a Poisson distribution:
\begin{equation}
p_{\Delta t}(k|n_k\varpi)=(1/k!)\,(x_{n_k})^k\,e^{-x_{n_k}};\qquad x_n\equiv
(\Delta t/\tau)\,2^{-n}.  \label{pk}
\end{equation}

Is this result not at variance with ones conception of black hole radiation as
thermal ?  For thermal radiation the number of quanta {\it in a given mode\/}
is distributed exponentially; here we have  a Poisson distribution.  However,
the discrepancy is merely a result of the different sort of question we ask
here: how is the number of quanta emitted during a time interval in all modes
of frequency $n_k\varpi$ distributed ?

To see that the Poisson distribution which answers this question is
consistent with a thermal distribution, consider a blackbody cavity  at
temperature $T$.  Let us look at {\it  all\/} quanta contained in a random
subvolume.  Each quantum drawn from that subvolume is subject to a Boltzmann
distribution; accordingly, the probability that we draw one quantum of
frequency $\omega$ is $A\exp(-\hbar\omega/T)$ where $A$ is a normalization
constant.  Succesive drawings are independent drawings from the same
distribution.  Thus the probability that we draw exactly $k$ quanta of
frequency $\omega_k$ from a series of $j$ drawings is $C^j_k
A^k\exp(-k\hbar\omega_k/T)\,A^{j-k}\Pi_{i=1}^{j-k}\exp(-\hbar\omega_i/T)$
summed over all possible values of the frequencies $\omega_i$ distinct from
$\omega_k$.  By normalization $A\sum_\omega \exp(-\hbar\omega/T)=1$, so if
the $\omega_k$ is missing from the sum, the corresponding factor in the
product, after the summation, will be $1-A\exp(-\hbar\omega_k/T)$.  We get
for $k$ a probability distribution exactly like Eq.~(\ref{intermediate})
with the replacement $2^{-n_k}\rightarrow A\exp(-\omega_k/T)$.  The perfect
analogy dispels the impression that the distribution for the black hole
radiance is not thermal.

 Another interesting question is what is the probability that $k_1$ quanta
are emitted into one group of modes, all of frequency  $n_k\varpi$, and $k_2$
into an equal number of modes of the same frequency (distinguished from the
first group, say, by angular momentum) ?  Since there are $C^{k_1+k_2}_{k_1}$
ways to divide $k_1+k_2$ quanta of frequency $n_k\varpi$ into one class with
$k_1$ and one with $k_2$, and a total of $2^{k_1+k_2}$ such bipartitions,
\begin{equation} p_{\Delta t}(k_1,k_2|n_k\varpi)=p_{\Delta
t}(k_1+k_2|n_k\varpi)\,2^{-(k_1+k_2)}\,C^{k_1+k_2}_{k_1}.
\end{equation}
With help of Eq.~(\ref{pk}) this takes the form
\begin{equation}
p_{\Delta t}(k_1,k_2|n_k\varpi)=p_{\Delta t/2}(k_1|n_k\varpi)p_{\Delta
t/2}(k_2|n_k\varpi)
\end{equation}
That is, the probabilities for $k_1$ and $k_2$ are independent and
of Poisson type, but their means are, naturally, reduced to half of that for
$k_1+k_2$.  The generalization to the probability for many sets of modes of
equal frequencies is obvious.

Let us now inquire into the probability $p_{\Delta t}
(k,l|n_k\varpi,n_l\varpi)$ for the emission of $k$ quanta of frequency $n_k$
and $l$ quanta of different frequency $n_l$ in time $\Delta t$.  There are
$j![k!l!(j-l-k)!]^{-1}$ ways for the $k$ and $l$ quanta to be
chosen---without regard to identity---from the $j$. The probability of each
such choice is $2^{-(n_1+n_2 +\cdots + n_{j-k})}(2^{-n_k})^k\,(2^{-n_l})^l$.
We sum the product over all allowed values of  $n_1\not= n_k$ and  $n_1\not=
n_l$, {\it etc.\/}  By the logic already invoked, the sum over $n_i$ of each
of the $j-k-l$ factors $2^{-n_i}$ is replaced by a factor
$1-2^{-n_k}-2^{-n_l}$. Instead of Eq.~(\ref{intermediate}) we get
\begin{equation}
p_{\Delta t}(k,l|n_k\varpi,n_l\varpi, j)={j!\over
k!l!(j-k-l)!}\left(1-{1\over 2^{n_k}}-{1\over 2^{n_k}}\right)^j\left({1\over
2^{n_k}}-1\right)^k \left({1\over 2^{n_l}}-1\right)^l\label{intermediate2}
\end{equation}
We now multiply by $p_{\Delta t}(j)$ of Eq.~(\ref{pj}), and
sum over all $j \ge k+l$ to get the result
\begin{equation}
p_{\Delta t}(k,l|n_k\varpi,n_l\varpi)=p_{\Delta t}(k|n_k\varpi)\,p_{\Delta
t}(l|n_l\varpi)
\end{equation}
Thus the probabilities for emission into the two frequencies are independent
and of Poisson type.  The generalization of this result to $n$ frequencies is
obvious.

We conclude by commenting on the shape of the spectrum.  The mean value of the
number of quanta emitted at a given frequency $\omega=n\varpi$ is, according
to Eq.~(\ref{pk}), proportional to $\Delta t$ and to $2^{-n}$.  Since
emissions at diverse frequencies are independent, it follows that the
intensity of the line at frequency $n\varpi$ must be proportional to $n
2^{-n}$ or equivalently to $\omega e^{-8\pi M\omega}$.  In accordance with our
earlier assumption, this is independent of particle species.  The spectrum is
to be compared with the Planckian spectrum (phase space factor aside)
$\omega/(e^{\hbar\omega/T}-1)$.  The black hole spectrum is obviously similar
to that of thermal radiation of like temperature in a finite box.

We found earlier that the rate of emission of quanta is $\tau^{-1}$.  Thus
the width of the lines in frequency should be of order $\tau^{-1}$.
But we concluded that $\tau$ is between a few hundreds and a few
tens of thousands of $M$.  We thus expect the width of the lines in frequency
to be small compared to $M^{-1}$, which is well below the size of the
frequencies, Eq.~(\ref{basicfreq}).  Therefore, the various
black hole lines are unlikely to overlap; the black hole spectrum must be
clearly discrete.  This conclusion differs from Hawking's original prediction
even for massive black holes.  Fluctuations of the horizon are here seen as
manifested in a discretization of the radiance's spectrum.  No quanta are
expected with frequencies below $M^{-1}$.  These  predictions will
become amenable to experimental check if primordial black holes are ever found.
They open a window upon quantum gravity that does not require reaching down to
Planck scale physics.

V. M. is grateful to the Racah Institute, Hebrew University, for hospitality.
J. D. B.'s work is partly supported by  a grant from the Israel Science
Foundation, administered by the Israel Academy of Sciences and Humanities.

\end{document}